\documentclass[aps,twocolumn,amsmath,amssymb,nobibnotes,superscriptaddress]{revtex4-1}

\usepackage{color}
\usepackage{amsmath}
\usepackage{amssymb}
\usepackage{graphicx}
\usepackage{epstopdf}
\usepackage[hidelinks,colorlinks=true,linkcolor=blue,citecolor=blue,anchorcolor=black,filecolor=cyan,menucolor=red,runcolor=cyan,urlcolor=magenta]{hyperref}
\usepackage{bm}



\begin{document}
\title{Theory of Domain-Wall Magnetoresistance in Metallic Antiferromagnets}
\author{Jun-Hui Zheng}
\affiliation{Center for Quantum Spintronics, Department of Physics, Norwegian University of Science and Technology, NO-7491 Trondheim, Norway}
\author{Arne Brataas}
\affiliation{Center for Quantum Spintronics, Department of Physics, Norwegian University of Science and Technology, NO-7491 Trondheim, Norway}
\author{Mathias Kl\"{a}ui}
\affiliation{Institute for Physics, Johannes Gutenberg-University Mainz, 55128 Mainz, Germany}
\affiliation{Center for Quantum Spintronics, Department of Physics, Norwegian University of Science and Technology, NO-7491 Trondheim, Norway}
\author{Alireza Qaiumzadeh}
\affiliation{Center for Quantum Spintronics, Department of Physics, Norwegian University of Science and Technology, NO-7491 Trondheim, Norway}
\date{\today}

\begin{abstract}
We develop a theory to compute the domain-wall magnetoresistance (DWMR) in antiferromagnetic (AFM) metals with different spin structures. In the diffusive transport regime, the DWMR can be either {\it negative} or positive depending on the domain-wall orientation and spin structure. In contrast, when the transport is in the ballistic regime, the DWMR is always positive, and the magnitude depends on the width and orientation of the domain wall. Our results pave the way of using electrical measurements for probing the internal spin structure in antiferromagnetic metals.
\end{abstract}

\maketitle
\section{Introduction}
Antiferromagnetic (AFM) materials are promising candidates for next-generation spintronic nanodevices with advantages of low-power consumption, fast spin dynamics, and small size \cite{Baltz2018,Nature2018}. In AFM memory devices, the data are stored in domains separated by domain walls (DWs) \cite{Foerster2014}. Highly efficient manipulation and accurate detection of AFM spin structures are essential for further developing the frontier of this field \cite{Baltz2018,Jungwirth2016}. The most extensively explored mechanisms for writing magnetic states in AFM metals so far are spin-transfer torques and spin-orbit torques \cite{Baltz2018,Jungwirth2016}. 
The magnetic state in a single domain can be electrically read out by measuring the anisotropic magnetoresistance (AMR) \cite{Marti2014}, the tunneling magnetoresistance (MR) \cite{Wang2014a} in AFM spin valves, and the tunneling  AMR \cite{Park2011,Duine2011}.

The comprehensive interplay between charge or spin currents and DWs brings diverse interesting phenomena and opens new opportunities for spintronic applications. The spin dynamics of DWs have been systemically studied in the presence of charge and spin currents in both ferromagnetic (FM) and AFM systems \cite{Berger,Tatara,zhang2004,Thiaville_2005,Tatara2006,Mougin_2007,Jung,TATARA2008213,Ryu, Marrows2005, BOULLE2011159, Aliev_2003, Yamanouchi2004,Yamanouchi,Yamanouchi1726,Kim2011,Hinzke,Shiino2016,Gomonay2016,Swaving2011,Qaiumzadeh2013, Qaiumzadeh2018,Qaiumzadeh2014,Tserkovnyak2014,Park2020,Khalili2020}. These studies greatly enrich the methods of electrically manipulating the magnetic texture. Conversely, how itinerant electrons scatter off a DW conveys magnetic texture information. Charge and spin currents scattering from FM DWs \cite{Cabrera1974,Tatara1997,Levy1997,Goldbart1998,Hoof1999,brataas1999,Ebels2000,PhysRevB63, Kent2001,Bauer2012,Bieren2013} and spin currents scattering  from  AFM DWs \cite{ross2019,Qaiumzadeh2018,Qaiumzadeh2014,Tserkovnyak2014,Shen2020} have also been explored extensively. Yet how charge currents are scattered by AFM DWs and the associated DWMR remain unexplored theoretically. Related studies could provide useful means for the detection of AFM domain structures.

In FM metals, a DW usually acts as an effective magnetic barrier and increases the MR \cite{Goldbart1998,Cabrera1974,Levy1997,Tatara1997,Hoof1999,brataas1999}. Negative DWMR might appear in special cases when
either the DW enhances the electron decoherence and reduces the weak localization in disordered systems \cite{Tatara1997} or the relaxation time is spin dependent \cite{Brataas19991}. AFM metals, on the other hand, have more complex magnetic textures than FM metals \cite{Lan2017}. Consequently, the AFM DWMR may exhibit more exotic properties \cite{Jaramillo2007}. Recently, an experimental study on charge transport in the AFM metal $\text{Mn}_2\text{Au}$ has reported a surprising AMR with the opposite sign to the typical AMR in FM systems and furthermore AFM DWMR signatures where found that are not understood on a theoretical level so far  \cite{2019Bodnar,Jourdan2018}. This clearly calls for a study to fill the gap of a missing theory of AFM DWMR to complete the understanding of DWMR in systems with different magnetic orderings and symmetries. 

\begin{figure}[t]
\includegraphics[width=0.98\columnwidth]{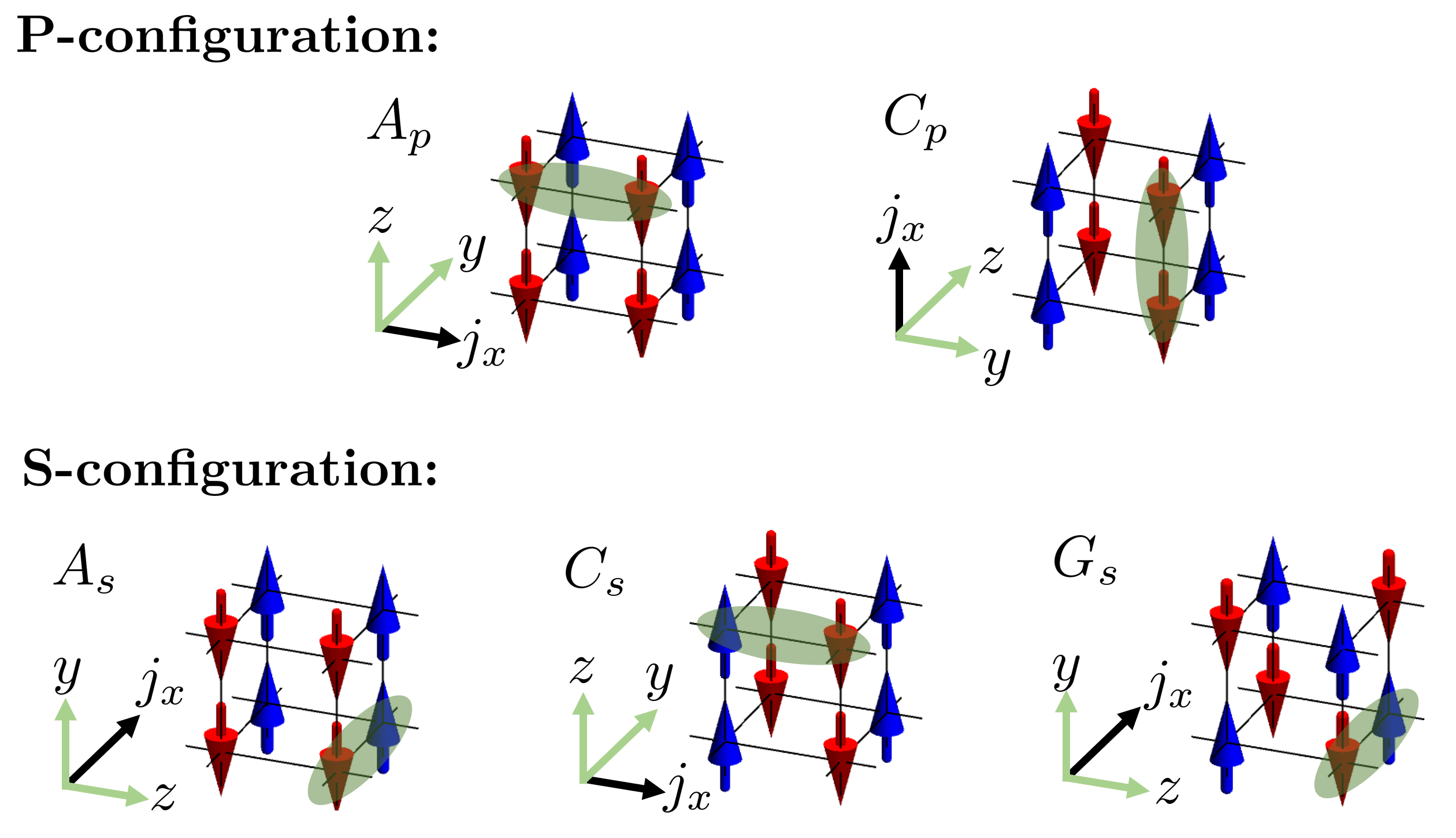}
\caption{A-type, C-type, and G-type AFM structures. The black vectors show the direction of the charge current $j_x$. In parallel (P) and staggered (S) configurations, spins are  parallel and antiparallel for neighbors in the direction of current, respectively. The DW is perpendicular to the current direction}
\label{tab}
\end{figure}

In this paper, we formulate a framework to explore the DWMR in AFM metals in cubic lattices with A-type, C-type, or G-type AFM spin structure as shown in Fig.\,\ref{tab}. Our formalism is general and can be used for other AFM lattices.
Assuming that the DW is pinned, we investigate the charge transport in the direction perpendicular to the DW. The local magnetic moments of the neighbors in the charge current direction may be ordered in parallel (ferromagnetically) or in antiparallel (antiferromagnetically). We name these two configurations as parallel (P) and staggered (S) configuration, respectively, see Fig.\,\ref{tab}.  Assuming that the Fermi wavelength $l_F$ is much smaller than the DW width $w$, we treat the transport problem in the diffusive regime and the ballistic regime, separately. The two regimes correspond to different cases that the mean free path of the itinerant electrons is significantly smaller than the DW width, $l_{\text{MFP}} \ll  w$, and oppositely $l_{\text{MFP}}  \gg  w$.

\begin{figure}[t]
\includegraphics[width=0.96\columnwidth]{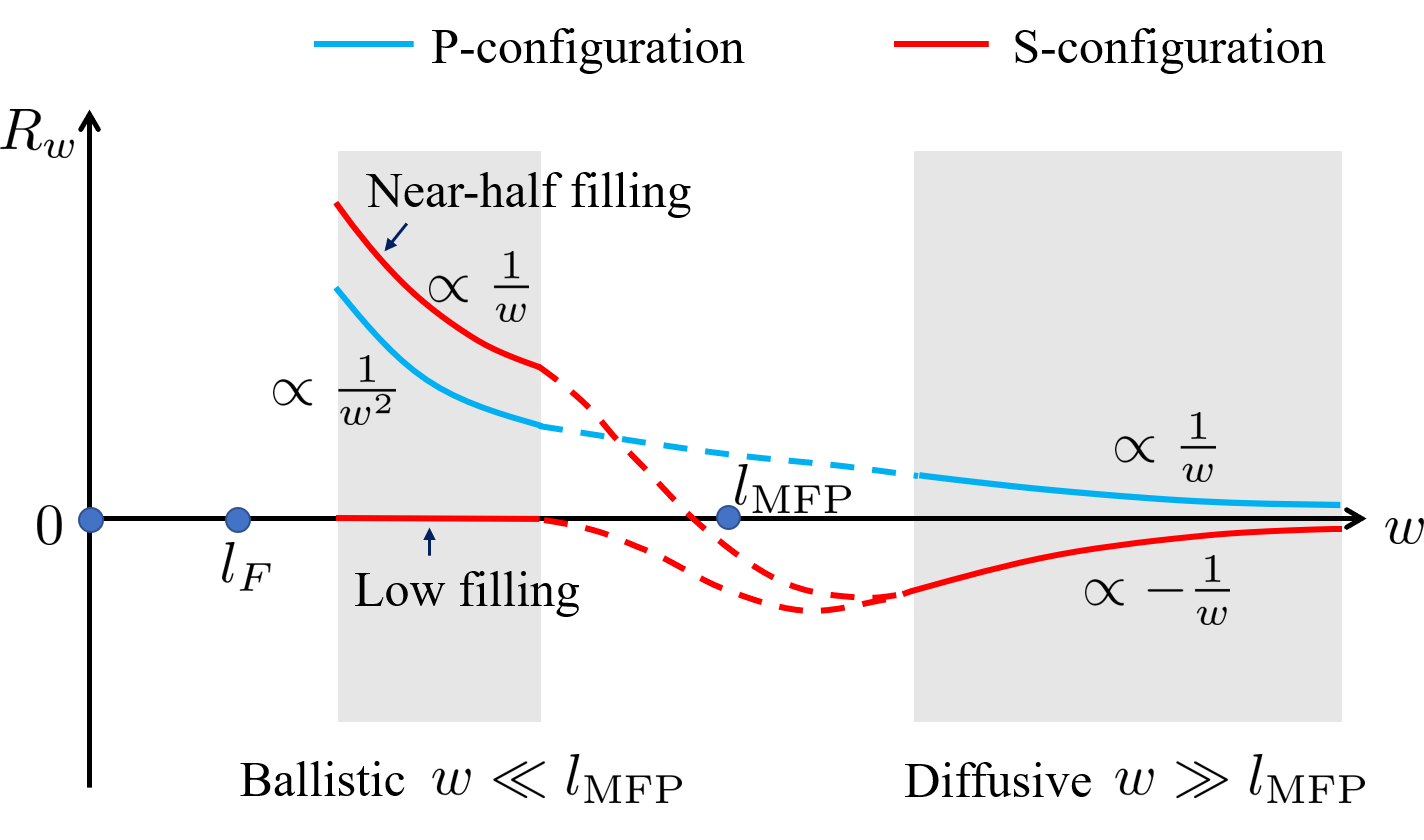}
\caption{Domain-wall magnetoresistance (DWMR) $R_w$ as a function of DW width $w$.  In P-configurations, the DWMR is positive. In S-configurations, it depends on the ratio between the domain wall width and the  mean free path of electrons $l_{\text{MFP}}$. In the diffusive regime, the DWMR is negative in most cases while in the ballistic regime the DWMR is positive. Here, $l_F$ is the Fermi wavevector.}
\label{dwmr}
\end{figure}

In Fig.\,\ref{dwmr}, we schematically present our main result. We find that the DWMR is always positive in the P-configurations. In addition, its magnitude is proportional to $ 1/w$ in the diffusive regime and $ 1/w^2$ in the ballistic regime. These behaviors are very similar to the DWMR in FM metals \cite{brataas1999}. 
In contrast, the DWMR in the S-configurations strongly depends on ratio between the DW width and the mean free path as well as doping level. In this case, the DWMR is negative in the diffusive regime for most doping levels. The DW promotes rather than hinders the electron mobility. This is because the DW can effectively suppress the magnetic staggering stiffness and thus enhance the electron mobility. On the other hand, in the ballistic regime, the DWMR becomes positive. It is proportional to $ 1/w$ near-half filling (about one electron per cubic cell) and vanishes in the low filling. Consequently, the DWMR changes its sign when the DW width becomes comparable to the mean free path for the intermediate electron filling case in the S-configurations.

The rest of this paper is structured  as follows. In Sec. \ref{secham}, we introduce the generic Hamiltonian and its symmetry. In Sec.\,\ref{dtt} and \ref{btt}, we develop the diffusive transport theory and the ballistic transport theory separately and calculate the DWMR. In Sec.\,\ref{concl}, we give a short conclusion.

\section{Hamiltonian}\label{secham}
We model the itinerant electrons in two-sublattice AFM metals, containing  antiferromagnetically coupled A and B sublattices, with the following generic Hamiltonian
\begin{equation}\label{eq:Hamiltonian}
    \mathcal{H}= -J \bm{n}(x)\cdot\bm{\sigma}
\tau_z + h_1(\bm{k}) \sigma_0  \tau_x +  h_2 (\bm{k})\sigma_0  \tau_0,
\end{equation}
where $\bm k$ is the electron wavevector and $J$ is the strength of the {\it{s-d}} exchange interaction between the itinerant electrons and the staggered local magnetization $\bm{n}$  \cite{Baltz2018,niu2012}. Without loss of generality, we consider a N{\'e}el-type DW perpendicular to the $x$ axis. The local magnetization is parameterized as \begin{equation}
    \bm{n}(x)=(\sin\theta \cos\phi, \sin\theta\sin\phi,\cos\theta), 
\end{equation}  
where $\phi$ is the azimuthal angle, which is a constant in N{\'e}el DWs,  and $\theta$ is the polar angle, which depends on the position along the $x$ direction. The Pauli matrices $\bm {\tau}$ and $\bm{\sigma}$ act on the two-sublattice subspace and the spin subspace, respectively. 
The structure factor 
\begin{equation}
h_1(\bm{k})=-t\sum_{\bm{d}}\cos{(\bm{k\cdot d})},  
\end{equation}  
describes inter-sublattice hopping between the antiferromagnetically ordered nearest-neighbor (A-B sites) with connection bonds $\bm d$. The structure factor \begin{equation}
h_2(\bm{k})=-t \sum_{\bm{b}}\cos{(\bm{k\cdot b})},  
\end{equation} 
governs intra-sublattice hopping between ferromagnetically ordered nearest-neighbor (A-A or B-B sites) with connection bonds $\bm b$, if such hoppings exists, e.g., for the G-type AFM structure, this term is zero. The explicit form of $h_1$ and $h_2$ for each AFM spin structure is shown in Table \ref{tbs}. 

\begin{table}
\caption{Hamiltonian $\mathcal H $ and the spin-spiral spectrum $E_\zeta^s$}
\label{tbs}
\begin{tabular}{|c|l|}
\hline
 Type
&$\begin{array}{l} \mathcal H = -J \bm{n}\cdot\bm{\sigma}
\tau_z + h_1(\bm{k}) \sigma_0  \tau_x +  h_2 (\bm{k})\sigma_0  \tau_0 \\
\text{and spectrum~} E_\zeta^s  \text{~with~} s,\zeta=\pm 1  \end{array}$
\tabularnewline
\hline
$A_s$
&  $\begin{array}{l}
h_1  =  -2t \cos k_x \\
h_2  = -2t (\cos k_y  +\cos k_z)  \\
E^s_\zeta =  -2t \cos k_y   -2t \cos k_z    \\
 ~~  + 2 \zeta  t\sin\frac{\lambda}{2}\sin k_{x} +s\sqrt{4t^{2}\cos^{2}\frac{\lambda}{2}\cos^{2}k_{x}+J^{2}}
 \end{array}$
\tabularnewline
\hline
$A_p$
&
$\begin{array}{l}   h_1  = -2t \cos k_y \\
  h_2  = -2t (\cos k_x + \cos k_z)   \\ 
E^s_\zeta = -2t\cos\frac{\lambda}{2}\cos k_{x}-2t\cos k_{z}\\
~~ +s\sqrt{4t^{2}(\cos k_{y}-\zeta \sin\frac{\lambda}{2}\sin k_{x})^{2}+J^{2}} \end{array} $
\tabularnewline
\hline
$C_s$
 &
$\begin{array}{l}   h_1  =  -2t( \cos k_x  + \cos k_y) \\
   h_2  =  -2t \cos k_z 
\\
E^s_\zeta = -2t \cos k_z + 2 \zeta  t\sin\frac{\lambda}{2}\sin k_{x}\\
 ~~+s\sqrt{4t^{2}[\cos\frac{\lambda}{2}\cos k_{x} + \cos k_y ]^2 +J^{2}}
 \end{array}$
\tabularnewline
\hline
$C_p$
&
$\begin{array}{l}   h_1  =  -2t( \cos k_y  + \cos k_z)   \\   
 h_2  =  -2t \cos k_x \\
E^s_\zeta = -2t\cos\frac{\lambda}{2}\cos k_{x}\\
~ +s\sqrt{4t^{2}(\cos k_{y}+\cos k_z-\zeta \sin\frac{\lambda}{2}\sin k_{x})^{2}+J^{2}} \end{array}$
  \tabularnewline
\hline
$G_s$  &
$\begin{array}{l}   h_1  =  -2t(\cos k_x + \cos k_y + \cos k_z) \\
 h_2  = 0  \\
E^s_\zeta = 2 \zeta  t\sin\frac{\lambda}{2}\sin k_{x}\\
~~+s\sqrt{4t^{2}[\cos\frac{\lambda}{2}\cos k_{x} + \cos k_y  +\cos k_z]^2 +J^{2}}
  \end{array}$
\tabularnewline
\hline
\end{tabular}
\end{table}

Since the Hamiltonian is spatially dependent along the $x$ direction, we replace the wavevector $\bm k$ with the  operator $\hat{\bm k}=({-i\partial_{x}}, k_y,k_z)$, where the components $k_y$ and $k_z$ remain good quantum numbers. Next, we apply a gauge transformation
\begin{equation}
    \mathcal{R}(x) = \exp{\left[- \frac{i \phi \sigma_z} {2}\right]}\exp{\left[- \frac{ i\theta (x) \sigma_y }{2}\right]},
\end{equation}
which makes the exchange term spatially uniform \cite{Tatara1997}, i.e., 
\begin{equation}
    \mathcal{R}^{-1}[\bm{n}(x) \cdot \bm{\sigma}]\mathcal{R}=\sigma_z.
\end{equation}
Simultaneously, 
the operator $-i\partial_{x}$ in the hopping terms $h_1$ and $h_2$ also becomes
\begin{equation}
   \mathcal{R}^{-1}[-i\partial_{x}]\mathcal{R}=-i\partial_{x}- \frac{\lambda(x)\sigma_{y}}{2},
\end{equation}
where $\lambda(x)=d\theta/dx$ describes the spatial gradient of the DW texture. The DW now induces a non-Abelian gauge potential $\lambda(x)\sigma_{y}/{2}$ inside the hopping terms, which vanishes far from the DW. To make the matrix representation of the Hamiltonian more elegant, we further apply a global rotation transformation
\begin{equation}
  \mathcal{T} = \exp{\left[-\frac{{i\sigma_{x}}\pi}{4}\right]}\exp{\left[-\frac{i\tau_{y}\pi}{4}\right]}.
\end{equation}
Finally, the Hamiltonian in the rotated basis becomes
\begin{eqnarray} \label{effective-hamiltonian}
{\mathcal{H}}_r &=& [\mathcal{R}(x) \mathcal{T}]^{-1}\mathcal{H}\mathcal{R}(x)\mathcal{T} \notag \\
&=& J  \sigma_y \tau_x + h_1(\hat{\bm k} + \frac{\lambda \sigma_z}{2} \bm {e}_x  ) \tau_z  +  h_{2}(\hat{\bm k} +  \frac{\lambda \sigma_z}{2}\bm {e}_x),
\end{eqnarray}
where $ {\bm{e}}_ x =(1,0,0)$. As well as the charge conservation, the rotated Hamiltonian has pseudospin conservation, since $[\zeta, {\mathcal{H}}_r]=0$,
where  $\zeta =\sigma_z\tau_z$ is the pseudospin operator. The matrix form of $\mathcal{H}_r$ is block diagonal. The spin-spiral spectrum of $\mathcal{H}_r$ with a constant $\lambda$ are shown in Table \ref{tbs}.

In the following, we compute the DWMR in both diffusive and ballistic regimes using this rotated Hamiltonian. The lattice constant is set to be $a=1$ for simplicity. We will assume that the DW texture is modelled by \cite{Qaiumzadeh2018,Qaiumzadeh2014}
\begin{equation}
  \cos\theta = \tanh\left(\frac{\pi x}{w}\right).
\end{equation}
Then, the gradient of the DW texture is maximum at the DW center, $|\lambda|_{\text{max}}=\pi/w$.

\section{ Diffusive transport theory}\label{dtt}
When the mean free path is significantly shorter than the DW width $w$, electrons move diffusively. The corresponding DWMR in FM systems has previously been evaluated using perturbative quantum field theory \cite{Tatara1997,brataas1999} and Boltzmann transport theory \cite{Levy1997}. To circumvent the complicated evaluation of Feynman diagrams in quantum field theory, we provide a new and considerably simpler method for computing the diffusive transport when the DW is wide $w \gg l_F$ and $\lambda(x)$ varies slowly. 

\begin{figure}[t]
\includegraphics[width=0.8\columnwidth]{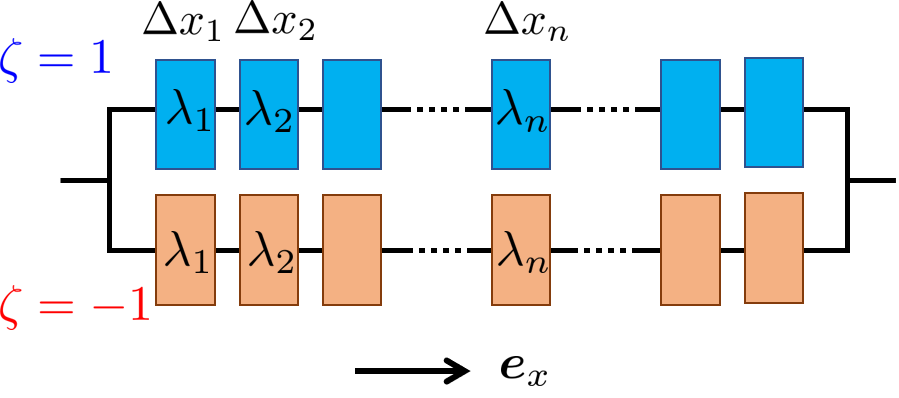}
\caption{The equivalent resistor circuit of an AFM system in the diffusive regime.}
\label{circus}
\end{figure}

As shown in Fig.\,\ref{circus}, we divide the system into a series of small spin-spiral segments with length $\Delta x_i$. Each segmemt has a constant spin-spiral gradient $\lambda_i = \lambda (\bar{x}_i)$, where $\bar{x}_i$ is the center position of the $i$-th segment. In the diffusive regime, Ohm's law applies. Since the pseudospin is conserved, the two pseudospin flavors $\zeta= \pm 1$ function as parallel resistor circuits. For each pseudospin flavor $\zeta$,
the resistance of a series resistor circuit is given by the sum of resistances of infinitesimal resistor elements
\begin{equation}
    R_\zeta= \sum_i \frac{\Delta x_i}{\sigma_\zeta(\lambda_i)S} = \int_{-L/2}^{L/2} \frac{dx }{\sigma_\zeta(\lambda) S},
\end{equation}
where $\sigma_\zeta(\lambda_i)$ is the conductivity of the $i$-th spin-spiral segment with pseudospin $\zeta$, $L$ is the system length along the $x$-axis, and $S$ is the cross-section. The total resistance of the system $R$ consists of the two pseudospin flavors in parallel, ${1}/{R}= {{1}/{R_+} + {1}/{R_-}}$. The DWMR is the difference between the total MR in the presence and in the absence of a DW, 
\begin{equation}\label{rw}
  R_{w}=R-R_{0}\simeq- \sigma_{0}^{-2}S^{-1} \int_{-L/2}^{L/2}\delta\sigma dx,
\end{equation}
where $R_0 = L/ S\sigma_0 $ is the MR of a single AFM domain, 
\begin{equation}
\sigma (\lambda) = \sigma_+(\lambda) + \sigma_-(\lambda) 
\end{equation}
is the total conductivity in a spin-spiral segment, $\sigma_0 = \sigma(0)$, and $\delta \sigma  =\sigma(\lambda) - \sigma_0$. In Eq.\,\eqref{rw}, we have used that in a uniform domain $\sigma_+(0) = \sigma_-(0)$. Now, the problem of DWMR has been reduced to the calculation of the conductivity $\sigma_\zeta(\lambda)$ in a spin-spiral segment. Note that depending on specific materials, the system can have different electron filling per cubic cell.  

In a low-filling regime, where only the low energy states are occupied, the chemical potential $\mu$ is close to the band bottom at $ - 6t$. The low-energy eigenvalues of ${\mathcal{H}}_r$ for a spin-spiral segment with a constant $\lambda$ becomes especially simple. From the spectrum shown in Table \ref{tbs}, up to the second order in $\lambda$ and $k_\alpha$, we find
\begin{equation}\label{spec}
    E \simeq  \delta E + \frac{(k_x - k_c)^2}{2m_x^*} +  \frac{k_y^2}{2m_y^*}+  \frac{k_z^2}{2m_z^* } + \text{const.}
\end{equation}
Here, the anisotropic effective mass of electrons is $m^*_\alpha = g_\kappa/2 t^2$ or $1/2 t$, when the spins are staggered or parallel for neighbors along the $\alpha$ axis. The numerator is defined as $g_\kappa =\sqrt{ t^2 +{J^2}/{4 \kappa^2}}$, where $\kappa=1, 2, 3$ for A-type, C-type, and G-type AFM structures, respectively.
A finite spin-spiral gradient leads to a shift in the momentum $k_c \propto \lambda$ and the energy $ \delta E = \mathcal{Z}_{\kappa} \lambda^2/m^*_x $, where $\mathcal{Z}_{\kappa} =  {J^2}/{(32 g_\kappa^2 \kappa^2)}$ for P-configurations and $\mathcal{Z}_{\kappa} =  - {J^2}/{(32 t^2 \kappa^2)}$ for S-configurations respectively. The spectrum \eqref{spec} is similar to the sepectrum of a free electron model and thus we can use the Drude formula for conductivity.
The Drude conductivity along the $x$ direction is given by 
\begin{equation}\label{Drude}
\sigma (\lambda) = e^2 n \tau/ |m_x^*|,
\end{equation}
where $n = k_F^x k_F^y k_F^z/3\pi^2$ is the electron density, $k_F^\alpha$ is the anisotropic Fermi wavevector, and $\tau$ is the electron lifetime. 
Using Eq.\,\eqref{rw}, we find that the relative change in the MR due to the DW becomes
\begin{equation}\label{rwr0}
\frac{R_w}{R_0} =  6 \pi \mathcal{Z}_\kappa \cdot \frac{ (l_F^x)^2}  { Lw},
\end{equation}
where $l^\alpha_{F} =1/k^\alpha_{F}$ is the anisotropic Fermi wavelength.

Equation \eqref{rwr0} shows that in  P-configurations,  $A_p$ and $C_p$ cases, the DWMR is positive similar to FM systems. In contrast, in S-configurations, $A_s$, $C_s$, and $G_s$ cases, the DWMR becomes negative, because the coefficient $\mathcal{Z}_\kappa$ is negative. The counter intuitive phenomena that these DWs reduce the MR, can be interpreted from the Drude's formula. In S-configurations, since the energy shift induced by a spin-spiral is negative, the local electron density and also the local conductivity inside the DW are enhanced. In other words, the DWs soften the effects of the staggered field.

Next, we consider a general case with arbitrary electron filling. Note that for a spin-spiral with a constant gradient $\lambda$, the momentum $k_x$ becomes a good quantum number in  ${\mathcal{H}}_r$. Using the Kubo formalism \cite{mah00}, we prove that the conductivity for each pseudospin becomes 
\begin{equation}\label{sigma}
\sigma_\zeta(\lambda) = {e^{2}}\tau \sum_{s}\int_{\text{BZ}}\frac{d^{3}\bm{k}}{(2\pi)^{3}}(v_{x}^{\zeta s})^{2}\delta(E_\zeta^s-\mu),
\end{equation}
where $v_{x}^{\zeta s} = \partial E_\zeta^s /\partial k_x$ is the group velocity of electrons in the $s$-th band of pseudospin $\zeta$. The relative DWMR can be shown to be ${R_w}/{R_0} \propto 1/w$.  Technical details can be found in the Appendix.
In the low-filling limit, Eq.\,\eqref{sigma} is consistent with the Drude conductivity \eqref{Drude}. We can also rewrite the conductivity \eqref{sigma} as $\sigma_\zeta(\lambda) = {e^{2}}\tau \overline{v_x^2} \rho_\zeta^\mu $, where $\rho_\zeta(\mu) $ is the density of states (DOS) and $\sqrt{\overline{v_x^2}}$ is the average velocity along the $x$-direction at the chemical potential $\mu$ for pseudospin $\zeta$. This expression clearly shows that the conductivity can be enhanced by increasing either the DOS or the average velocity.

\begin{figure}[t]
\includegraphics[width=1.0\columnwidth]{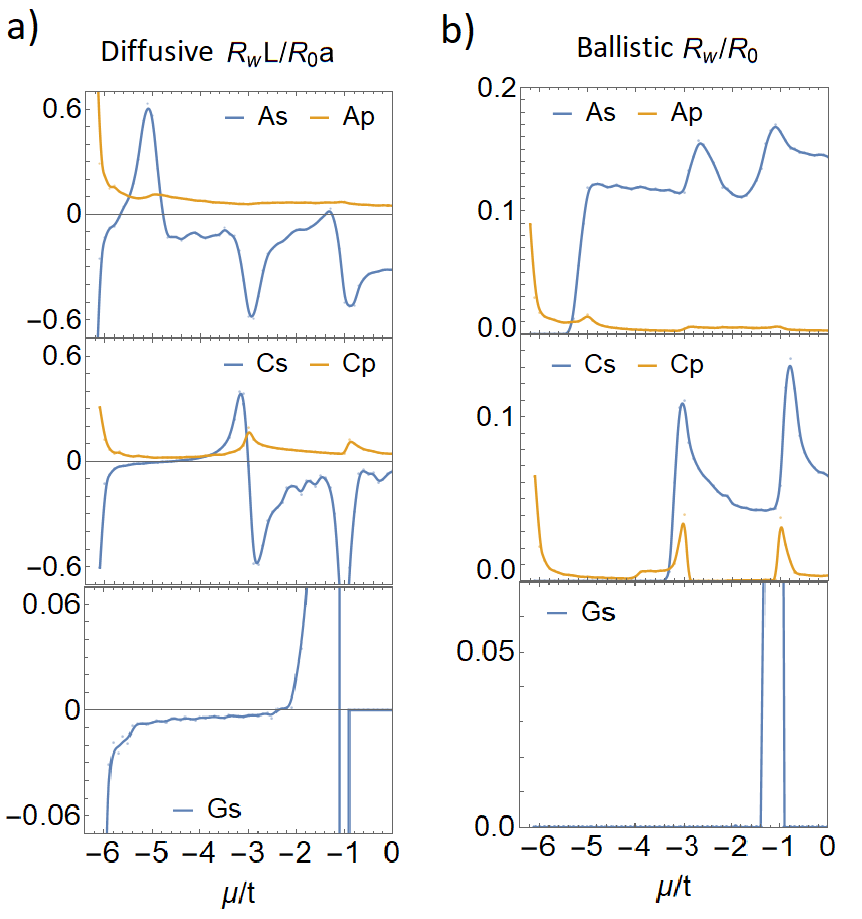}
\caption{ The relative DWMR $R_w/R_0$ as a function of the chemical potential $\mu$. The DW width is $w\simeq 10.5 a$ and we set $J=t$. We show the results for negative chemical potential. The part with positive chemical potential is symmetric with the negative part. Here, $L$ is the length of the system and $a$ is the lattice constant.}
\label{sigmas}
\end{figure}

\begin{figure}[t]
\includegraphics[width=\columnwidth]{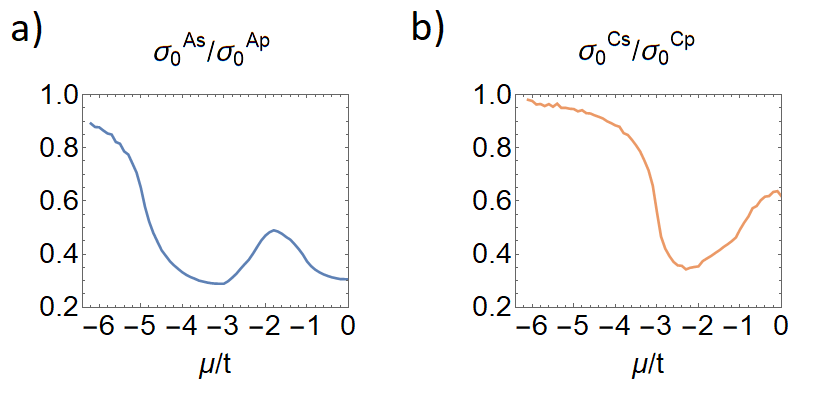}
\caption{ The ratio of diffusive conductivity in a single AFM domain with S-configuration and P-configuration. The ratio is significantly less than one.}
\label{tab3}

\end{figure}

\begin{figure}[t]
\includegraphics[width=\columnwidth]{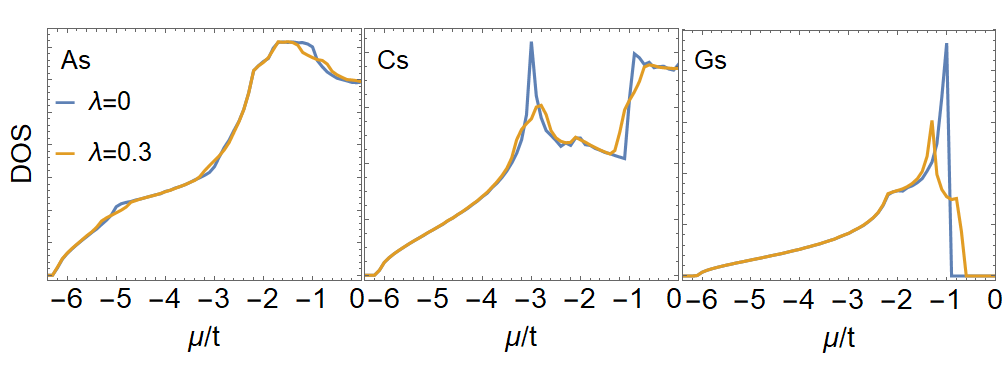}
\caption{Density of states (DOS) for $A_s$, $C_s$ and $G_s$ cases}
\label{tab4}
\end{figure}

In Fig.\,\ref{sigmas}-a), we plot the relative DWMR as a function of chemical potential for a DW with $\lambda_{\text{max}} =0.3/a$. The DW width is $w \simeq 10.5 a$. The result shows that for different chemical potentials, the DWMR usually is negative in S-configurations and always is positive in P-configurationa. To understand this phenomena, in Fig.\,\ref{tab3}, we confirm that the ratios of the conductivity of uniform domains ($\lambda=0$) for different AFM types, $ \sigma_{0}^{{ A}s}/\sigma_{0}^{{ A}p}$ and  $ \sigma_{0}^{{C}s}/\sigma_{0}^{{ C}p}$, are significantly smaller than one. Since the single AFM domains of $A_s$-type and $A_p$-type, or $C_s$-type and $C_p$-type, share a common DOS, the ratio less than one means that the electron mobility (or the averaged velocity) is suppressed in the direction that neighboring magnetic moments are staggered. We interpret that in S-configurations a spin spiral can effectively suppress the magnetic staggering stiffness and thus enhances the electron mobility. This explains the formation of negative DWMR in S-configurations. In Fig.\,\ref{sigmas}-a), we also find in several energy regions in S-configurations, the DWMR becomes positive. These phenomena usually accompany with significant suppression of DOS (corresponding to the conducting channels) due to the spin spiral as shown in Fig.\,\ref{tab4}.


\section{Ballistic transport theory}\label{btt}
When the electron's mean free path is much larger than the DW width, by using the Landauer approach, the conductance in the $x$-direction at zero temperature is
\begin{equation} \label{bt}
G= {e^{2}}S \sum_{\zeta s}\int_{\text{BZ}}^{v_{x}^{\zeta s}>0}\frac{d^{3}\bm{k}}{(2\pi)^{3}}v_{x}^{\zeta s}|_{\lambda=0} T_{\zeta s,k_{x}}\delta(E_\zeta^s|_{\lambda =0} -\mu),
\end{equation}
where $T_{\zeta s,k_{x}}$ is the transmission coefficient for each channel \cite{nazarov_blanter_2009}. 
Since $\lambda(x)$ varies slowly, the adiabatic approximation can be applied. In other words, we can treat the motion of electrons classically by using Hamilton's equations, $\dot{x}=\partial_{k_{x}}E_{\zeta}^{s}(\lambda(x),k_x)$ and $\dot{k}_{x}=-\partial_{x}E_{\zeta}^{s}(\lambda(x),k_x)$. The two variables $k_y$ and $k_z$ are hidden since they are conserved during the evolution.
The transmission coefficient becomes
$T_{k_{x}}=1$ for open channels and $T_{k_{x}}=0$ for closed channels. The DWMR is determined by how many  channels
become closed due to the DW.
Note that the electrons' motion follows the law of conservation of energy. The DW, as an aisle, only allows injecting electrons within the energy window $\big[\min(E_\zeta^s),~\max(E_\zeta^s)\big]_{\lambda=\lambda_{\text{max}}}$ for fixed $k_y$ and $k_z$ set by the DW to pass through. Technically, for an injecting electron with $E=E_{\zeta}^{s}(\lambda=0,k_x^0)$, the channel is open if there is a solution of $k_x$ near to $k_x^0$ for the equation $E=E_{\zeta}^{s}(\lambda_{\text{max}},k_x)$. 

In the low filling limit, using the low energy spectrum \eqref{spec}, we obtain the resistance of a single AFM domain, $R_0  = ({4\pi^2 \hbar}/ {e^{2}}) ({l_F^y l_F^z}/{S}) $.
For all S-configurations, the DWs are transparent and cause no additional resistance, since all low-energy channels are open due to the negative energy shift $\delta E $. However, for P-configurations, $\delta E$ is positive, we obtain the relative DWMR
\begin{equation}\label{rwr0x}
 \frac{R_w}{R_0} 
= \frac{(\pi l_F^x)^2}   {w^2} \cdot 2\mathcal{Z}_\kappa.
\end{equation}

The situation differs in the moderate filling case. For instance, in the $A_s$ case, the local dispersion with gradient $\lambda$ is 
\begin{eqnarray}
E^\pm_\zeta &=& -2t(\cos k_y + \cos k_z) +   2 \zeta  t\sin({\lambda}/{2})\sin k_{x} \notag\\
&& \pm \sqrt{4t^{2}\cos^{2}({\lambda}/{2})\cos^{2}k_{x}+J^{2}}.
\end{eqnarray}
We rewrite the energy in the form 
\begin{equation}
{E}^\pm_\zeta (\lambda,k_x) = {E}^\pm_\zeta(\lambda=0, k_x) + V_{\text{eff}}[\lambda(x),k_x].
\end{equation}
Different models can use a similar decomposition. Here we find near $k_x=\pi/2$ (which can not be occupied in the low filling limit), the effective potential  $V_{\text{eff}} \sim \pm 2 t \lambda(x)/2$ for a small $\lambda$. This quantity tunes the energy window in the DW. In S-configurations, the leading order of tuning the window is $\propto\lambda_{\text{max}}$ and thus $\delta{R_{r}} \propto {1}/{w}$. Similar discussion works for P-configurations. However, the leading order of tuning the window becomes proportional to $\lambda_{\text{max}}^2$, resulting in $\delta{R_{r}} \propto {1}/{w^2}$ as the low-filling case. 

In Fig.\,\ref{sigmas}-b), we plot the relative DWMR for a $180^\circ$ DW with $\lambda_{\text{max}} = 0.3/a$. We see in S-configurations it vanishes for a low filling case, but becomes significantly larger than that in P-configurations in a moderate filling case.

\section{Concluding remarks}\label{concl}
We have developed a formalism to compute the MR from magnetic textures in AFM metals in diffusive and ballistic regimes. In AFM systems, the DWMR exhibits a more complex behavior as compared to FMs. The AFM DWMR can be negative in the diffusive regime even in the absence of spin-dependent relaxation time and spin-orbit couplings, which contrasts the FM case. This unique feature arises because DWs soften the effects of the staggered magnetic moments in AFM metals.

Our results stimulate further DWMR measurements to reveal the properties of AFM metals. In the diffusive regime, since the sign of DWMR is sensitive to spin configurations, measuring the MR provides information on the configuration along the current direction. The qualitative dependence of the DWMR on DW properties is measurable by varying the DW width through mechanical strains or magnetic fields in conjunction with magnetic imaging using X-ray magnetic linear dichroism contrast \cite{ross2019,2019Bodnar}.

\begin{acknowledgements}
The authors thank O. Gomonay and M. Jourdan for discussions.
This work was supported by the European Research Council via an Advanced Grant (no.\,669442 ``Insulatronics''), the Research Council of Norway through its Centres of Excellence funding scheme (project no.\,262633, ``QuSpin''), the Norwegian Financial Mechanism 2014-2021 (Project No. 2019/34/H/ST3/00515, ``2Dtronics''), and the German Research Foundation (SFB TRR 173 Spin+X, projects A01 and B02). 
\end{acknowledgements}

\appendix

\appendix

\section{ Charge conductivity in the spin-spiral segment} 

The pseudospin $\zeta$ is a conserved in our systems. Below we focus on each subspace with $\zeta = \pm 1$. For the simplicity of notation, we hide the index $\zeta$ in the following. For a spin-spiral segment with a constant gradient $\lambda$, the momentum $k_x$ in the rotated Hamiltonian $\mathcal{H}_r$ becomes a good quantum number. Correspondingly, the current operator in the rotated basis $(\hat{c}^\dagger_{\bm k,s})$ is
\begin{equation}
\hat{j}_x = \sum_{\bm k, s,s'} \hat{c}^\dagger_{\bm k,s} {\mathcal{J}}_{ ss'}(\bm k)  \hat{c}_{\bm k, s'},
\end{equation}
where ${\mathcal{J}}(\bm k) = \partial_{k_x} \mathcal{H}_r(\bm k)$ and $s$ represents the internal degree of freedom besides the pseudospin $\zeta$.

Next, we introduce a unitary transformation $U_{\bm k}$ to diagonalize the Hamiltonian, ${\mathcal{H}}_{r}(\bm k) = U_{\bm k}^\dagger \Lambda_{\bm k} U_{\bm k} $, where $\Lambda_{\bm k} = \text{diag} \{ E^+(\bm k), E^-(\bm k) \}$ is a diagonal matrix. In the eigenbasis of the Hamiltonian, $  \hat{d}_{\bm k} = U_{\bm k}  \hat{c}_{\bm k}$, the current operator becomes
\begin{equation}
\hat{j}_x =  \sum_{\bm k, s,s'} \hat{d}^\dagger_{\bm k,s} J_{\bm k; ss'}  \hat{d}_{\bm k, s'},
\end{equation}
where  $J_{\bm k}= U_{\bm k} {\mathcal{J}}(\bm k) U_{\bm k}^\dagger$.

The conductivity in the Kubo formalism is \cite{mah00}
\begin{equation}
\sigma = \lim_{\omega  \rightarrow 0}   \Big\{ \frac{\text{Im}[\pi(i \omega_n \rightarrow \omega + i\delta)]}{\omega}\Big\},
\end{equation}
where the current-current correlation is
\begin{eqnarray}
\pi(i\omega_n) &=& \frac{1}{V}\int_0^\beta dt e^{i\omega_n t} \langle T_t \hat{j}_x(t)\hat{j}_x(0)\rangle \notag\\
&=& \frac{1}{ \beta} \sum_{p}\int \frac{d^3\bm k}{(2\pi)^3} \text{Tr}\big[ {J}_{\bm k}
\mathcal{G}_{\bm k}(ip + i\omega) {J}_{\bm k} \mathcal{G}_{\bm k}(ip)
\big].\notag\\
\end{eqnarray}
The single particle Green's function $ \mathcal{G}_{\bm k}(ip)$ is a diagonal matrix. In the Lehmann representation, we have 
\begin{equation}
 \mathcal{G}_{\bm k;s,s'}(ip_n) =\delta_{s,s'} \int \frac{d\epsilon }{2\pi} \frac{A_s(\bm k, \epsilon)}{ip_n -\epsilon+\mu},
\end{equation}
where the spectral density is 
\begin{equation}
A_s(\bm k, \epsilon) =  \frac{ 2 \Delta_{\bm k} } {(E^s - \epsilon)^2 +\Delta_{\bm k}^2}.
\end{equation}
In above, $s=\pm$,  $\Delta_{\bm k} = 1/2\tau_{\bm k}$, where $\tau_{\bm k}$ is the lifetime of the quasiparticle.

Using these formula and using the same techniques developed in Section 8.1 of Mahan's book \cite{mah00}, we can directly obtain the conductivity,
\begin{equation}
\sigma =\sum_{s,s'}  \int \frac{d^3\bm k}{(2\pi)^3}  \frac{d \epsilon}{4\pi} {J}_{\bm k; s,s'}  {J}_{\bm k; s',s}  {A_s(\bm k, \epsilon)}{A_{s'}(\bm k, \epsilon)} \delta(\epsilon -\mu).
\end{equation}
Note that for a large $\tau_{\bm k}$ (i.e., a small $\Delta_{\bm k}$), we have 
\begin{equation}
    A_s(\bm k, \epsilon) A_{s'}(\bm k, \epsilon)\simeq 4 \pi \delta (\epsilon - E^s) \tau_{\bm k},
\end{equation} 
when $ E^s = E^{s'}$ \cite{mah00}, and
\begin{equation}
    A_s(\bm k, \epsilon) A_{s'}(\bm k, \epsilon) = 0,
\end{equation} 
when $ E^s \neq E^{s'}$.
On the other hand, using the fact that
\begin{equation}
   J_{\bm k} =  U_{\bm k} {\mathcal{J}}(\bm k) U_{\bm k}^\dagger = U_{\bm k} [\partial_{k_x} (U_{\bm k}^\dagger  \Lambda_{\bm k}  U_{\bm k} )] U_{\bm k}^\dagger,
\end{equation}
we obtain
\begin{equation}
  J_{\bm k}
=   U_{\bm k} ( \partial_{k_x} U_{\bm k}^\dagger) \Lambda_{\bm k}+ \partial_{k_x} \Lambda_{\bm k} +
\Lambda_{\bm k} ( \partial_{k_x} U_{\bm k})  U_{\bm k}^\dagger.
\end{equation}
Thus, for $ E^s = E^{s'}$,  we have
\begin{eqnarray}
J_{\bm k,s,s'} &=&   [ U_{\bm k} ( \partial_{k_x} U_{\bm k}^\dagger) +  ( \partial_{k_x} U_{\bm k})  U_{\bm k}^\dagger]_{s,s'}  E^{s'}_{\bm k}+  \delta_{s,s'}\partial_{k_x} E^{s'}_{\bm k} \notag\\
&=& \delta_{s,s'}\partial_{k_x} E^{s}_{\bm k}.
\end{eqnarray}
We further assume that $\tau_{\bm k} = \tau$ and finally obtain
\begin{equation}\label{sigmaxx}
\sigma = {e^{2}}\tau \sum_{s} \int \frac{d^{3}\bm{k}}{(2\pi)^{3}}(v_{x}^{ s})^{2}\delta(E^s-\mu),
\end{equation}
where $v_{x}^{ s} = \partial_{k_x}  E^{s}_{\bm k}$. We have added the factor $e^2$, which is set to $1$ in the above derivation.

By expanding Eq.\,\eqref{sigmaxx} as a function of the gradient $\lambda$, the correction from spin-spiral is of the order of $\lambda^2$.
 The relative DWMR becomes
\begin{equation}
\frac{R_{w}}{R_0}= \frac{\pi\mathcal{C}_{\text{diffuse}}}{Lw}, \label{rwxx}
\end{equation}
where
\begin{equation}\label{cdi}
\mathcal{C}_{\text{diffuse}}= - \frac{e^{2}\tau}{\sigma_{0}} \sum_{\zeta,s} \int_{\text{BZ}}\frac{d^{3}\bm{k}}{(2\pi)^{3}}  \mathcal{F}_2[E_{\zeta}^s] \delta(E_\zeta^s|_{\lambda =0}-\mu),
\end{equation}
and
\begin{eqnarray}
\mathcal{F}_2[E] &=&  \Big[    \ddot{E}' E'   -  2\dot{E} \dot{E}'' - \ddot{E}{E}'' + \frac{{E}'''}{E'}\dot{E}^2 \notag\\&& - \frac{{{E}''}^2}{{E'}^2}\dot{E}^2 
+ \frac{2{E}''}{E'}\dot{E}\dot{E}' \Big] \Big|_{\lambda =0}.
\end{eqnarray}
We have used the convention that $\dot{E} = \partial_\lambda E$ and ${E}' = \partial_{k_x} E$.  Eq.\,\eqref{rwxx} demonstrates that
in the diffusive transport regime, the DWMR is inversely proportional to the DW width. The following calculation is the detail for the expansion.

\section{Expansion of Eq.\,\eqref{sigmaxx}}

We start from the integral
\begin{equation}\label{int}
\mathcal{I} = \int_{\text{BZ}}\frac{d^{3}\bm{k}}{(2\pi)^{3}} v^{2}\delta(E-\mu),
\end{equation}
where $E(\bm k, \lambda)$ and $v(\bm k, \lambda) = E' =\partial_{k_x} E $ are functions of $\bm k$ and $\lambda$. Using the convention $\dot E = \partial_{\lambda} E $, we expand the velocity $v$ and the $\delta$-function around $\lambda =0$,
\begin{equation}
v^{2}\delta(E-\mu) =  \big(v_0 + \dot v_0 \lambda + \frac{1}{2}\ddot v_0 \lambda^2\big)^2  \big(\delta_0 + \dot \delta_0 \lambda + \frac{1}{2}\ddot \delta_0 \lambda^2\big),
\end{equation}
where $v_0 = v|_{\lambda =0}$ and $\delta_0 = \delta(E|_{\lambda =0}-\mu) $.
Order by order expanding the function $v^{2}\delta(E-\mu)$, we obtain coefficients for $\lambda^i$ as following,
\begin{subequations}
\begin{align}
\lambda^0: ~~~~~~~~~ & v_0^2 \delta_0  \\
\lambda^1: ~~~~~~~~~ &  2 v_0 \dot v_0 \delta_0 + v_0^2 \dot \delta_0 \\
\lambda^2: ~~~~~~~~~ &    \dot v_0 \dot v_0  \delta_0  + v_0 \ddot v_0  \delta_0 + \frac{1}{2} v_0 v_0 \ddot \delta_0 + 2 v_0  \dot v_0  \dot \delta_0. \label{coef}
\end{align}
\end{subequations}
For simplicity, we will omit the foot index $0$ but keep in mind $\lambda\rightarrow 0$.
Note that $\dot \delta = \dot E \partial_E \delta =    { \dot E} \delta'/v$ and
\begin{eqnarray}
\ddot \delta & = & \frac{\partial}{\partial\lambda}{( \dot E \partial_E \delta )} =    \ddot E \partial_E \delta  + \dot E {\frac{\partial}{\partial\lambda} ( \partial_E \delta )} \notag\\ & = & \ddot E \partial_E \delta  + \dot E ^2 {( \partial^2_E \delta )}
 = \frac{ \ddot E}{v} \delta'  + \dot E ^2  \frac{ 1}{v}{\frac{\partial}{\partial k_x}\Big(\frac{ 1}{v}  \delta'\Big )}  \notag\\ & = &   \frac{ \ddot E}{v} \delta' + \dot E ^2  \frac{ 1}{v^2}{ \delta''} - \dot E ^2  \frac{ v'}{v^3} \delta'.
\end{eqnarray}
Substituting these expansions into the integral \eqref{int}, Integration by parts shows that these coefficients become
\begin{eqnarray}
\lambda^1:&~~~~&  \big[ (E' \dot{E}' - E'' \dot{E}) \big]\big|_{\lambda =0} \delta(E|_{\lambda =0}-\mu),\\
\lambda^2: &&     \Big[\frac{1}{2}    \ddot{E}' E'    -  \dot{E} \dot{E}'' -\frac{1}{2}\ddot{E}{E}'' +\frac{1}{2}\frac{{E}'''}{E'}\dot{E}^2 \notag\\&&   -\frac{1}{2}\frac{{{E}''}^2}{{E'}^2}\dot{E}^2 + \frac{{E}''}{E'}\dot{E}\dot{E}' \Big] \Big|_{\lambda =0} \delta(E|_{\lambda =0}-\mu).
\end{eqnarray}
As a result, for both $\zeta$, we obtain
\begin{eqnarray}
\delta\sigma &=& e^2\tau \sum_{\zeta,s}\int_{\text{BZ}}\frac{d^{3}\bm{k}}{(2\pi)^{3}} \big\{
 \lambda   \mathcal{F}_1[E_{\zeta}^s]  \notag\\&& +  \frac{\lambda^2}{2} \mathcal{F}_2[E_{\zeta}^s]
\big\} \delta(E_\zeta^s|_{\lambda =0}-\mu),
\end{eqnarray}
where $
\mathcal{F}_1[E] = \big[E' \dot{E}' - E'' \dot{E} \big]\big|_{\lambda =0}.$
In our models, the parity symmetry of the functions under $k_x \rightarrow -k_x$, determines that the integral of $\mathcal{F}_1$ vanishes. Thus, the DWMR becomes
\begin{equation}
R_{w}=\frac{\mathcal{C}_{\text{diffuse}}}{\sigma_{0}S}\int_{-L/2}^{L/2}\frac{\lambda^2}{2} dx = \frac{\pi}{w}  \frac{\mathcal{C}_{\text{diffuse}}}{\sigma_{0}S}. \label{rwxxx}
\end{equation}
 Using $R_0 = L/\sigma_{0} S$, we finally obtain Eq.\,\eqref{rwxx}.


%

\end{document}